\documentclass[revsymb,bibnotes,amsmath,amssymb,reprint,superscriptaddress,floatfix]{revtex4-1}

\usepackage[]{graphicx}
\usepackage{bm}
\usepackage{float}
\usepackage{lineno}
\usepackage{hyperref}
\usepackage[all]{hypcap}
\usepackage{chngcntr}

\usepackage{lipsum}
\usepackage{bm}
\usepackage{slashed}
\usepackage{amsmath}
\usepackage{amsfonts}
\usepackage{amssymb}%
\usepackage{balance}

\makeatletter
\newcommand{\colorcaption}[2][]{%
  \begingroup%
  \renewcommand{\@caption@fignum@sep}{ (Color online). }%
  \caption[#1]{#2}%
  \endgroup%
}
\makeatother

\begin{document}

\author{F. J. T. Goncalves}\email{f-goncalves@pe.osakafu-u.ac.jp}
\affiliation{Department of Physics and Electronics, Osaka Prefecture University, 1-1 Gakuencho, Sakai, Osaka 599-8531, Japan}

\author{Y. Shimamoto}
\affiliation{Department of Physics and Electronics, Osaka Prefecture University, 1-1 Gakuencho, Sakai, Osaka 599-8531, Japan}

\author{T. Sogo}
\affiliation{Department of Physics and Electronics, Osaka Prefecture University, 1-1 Gakuencho, Sakai, Osaka 599-8531, Japan}
\author{G. W. Paterson}
\affiliation{SUPA, School of Physics and Astronomy, University of Glasgow, Glasgow, G12 8QQ United Kingdom}
\author{Y. Kousaka}
\affiliation{Graduate School of Science, Okayama University, Okayama 700-8530, Japan}

\author{Y. Togawa}
\affiliation{Department of Physics and Electronics, Osaka Prefecture University, 1-1 Gakuencho, Sakai, Osaka 599-8531, Japan}

\date{\today}
\title{Effect of disorder on the collective excitations of the chiral spin soliton lattice}

\date{\today}

\begin{abstract}
We assess the impact of magnetic disorder on the spin excitation spectra of the chiral helimagnetic crystal $\mathrm{CrNb_3S_6}$ using microwave resonance spectroscopy. The chiral spin soliton lattice phase (CSL), which is a prototype of a noncollinear spin system that forms periodically over a macroscopic length scale, exhibited three resonance modes over a wide frequency range. We found that the predominance of these modes depended on the degree of magnetic disorder and that disorder can be suppressed by sweeping the external field via an ideal helical state at 0~T. The macroscopic coherence of the CSL can be clearly observed through its collective dynamic behavior. Our study suggests that magnetic disorder can be used as a mechanism to control spin wave excitation in noncollinear spin systems.
\end{abstract}   
\maketitle
Recently, magnetic systems with noncollinear spin textures have attracted immense attention due to their ability to act as magnonic conduits with programmable band structure~\cite{Krawczyk2014,Davies2015,Davies2015a,Schwarze2015}.  
Naturally, good field stability and spatial coherence of the underlying noncollinear magnetic structure are essential requirements to ensure efficient propagation and manipulation of the magnonic signals. In this sense, chiral helimagnetic materials are very promising as the spatial coherence is built-in primarily due to an energy balance between the antisymmetric Dzyaloshinskii-Moriya (DM) interaction~\cite{Moriya1960,Moriya1982} and the symmetric exchange interaction, enabling spatially robust magnetic textures such as magnetic skyrmions~\cite{Schwarze2015,Garst2017,Weiler2017}, and the chiral helical and the chiral spin soliton lattice phases~\cite{Togawa2016}. In chiral helimagnetic materials, periodically self-assembled magnetic elements emerge and can be tuned efficiently with an external magnetic field. When dealing with magnetic materials capable of hosting these `naturally' assembled spin textures, there is the need to assess the impact of magnetic disorder and defects as these can disrupt the spatial coherence and consequently affect the collective excitations and the propagation of spin waves in the material.\par
In this letter, we shed light on the differences between the spin excitation spectra of the disordered and the ordered chiral spin soliton lattice phase of the monoaxial chiral helimagnetic crystal~$\mathrm{CrNb_3S_6}$. In micron-sized crystals, an energy barrier opposes the otherwise continuous transition from the field polarized to the CSL phase \cite{Shinozaki2018} and the disordered phase is obtained in the process of overcoming this energy barrier. We demonstrate how to control disorder and how to achieve an ordered CSL phase by use of a specific magnetic field protocol.\par
 
Microwave spectroscopy experiments on a micrometer sized crystal detected three resonance modes at the onset of the CSL phase. The lowest order modes were detected at 14-20~GHz while a higher order mode appeared at approximately twice this frequency. The multi-mode character of the spin excitation spectra is linked to the periodic nonlinear modulation of the moments forming the CSL and is reported here for the first time since its theoretical prediction~\cite{Kishine2009, Kishine2015, Kishine2016}. With further decreasing the field strength towards 0~T, we found that the resonance spectra at lower frequencies were comprised of two types of modes which we ascribe to intrinsic CSL modes and ferromagnetic, Kittel-like modes, indicating the existence of a disordered CSL phase. Interestingly, upon switching the polarization of external magnetic field through 0~T, we found a clear transformation in the amplitude and field dependence of the resonance modes measured in the increasing field branch towards the critical field. The emergence of the helical state at 0~T triggered the extinction of the disordered phase, enabling the recovery of the collective spin excitation expected for an ideal, ordered CSL phase. Thus, clear experimental evidence is provided on how the spin excitation spectra is particularly sensitive to the degree of disorder in the CSL of micrometer sized specimens.\par %

The spin configuration of the helimagnetic compound $\mathrm{CrNb_3S_6}$ at 0~T (below a critical temperature of 128~K) corresponds to the chiral helical state consisting of $\mathrm{2\pi}$ magnetic kinks (MKs) arrayed with a periodicity of 48~nm~\cite{Moriya1982, Togawa2012}. When a magnetic field is applied perpendicular to the helical axis, the chiral spin soliton lattice (CSL) emerges~\cite{Togawa2012}. The CSL is comprised of MKs periodically distanced by regions with field polarised spins. The periodicity of the MKs increases with the strength of $H$ until a critical field, $H_C$, is reached. Above $H_C$ a field polarised (FP) phase is obtained. The magnitude of $H_C$ is primarily defined by the DM and the exchange interaction constants and varies between 0.15 and 0.2~T~\cite{Togawa2015,Tsuruta2016b,Tsuruta2016a}. Demagnetization effects due to sample geometry and external field direction can induce variations in the magnitude of $H_C$~\cite{Goncalves2017}. Several interesting features of the CSL such as spatial coherence, robustness with regards to the magnetic field and discretized behavior can be found in the literature, both in the limits of small micrometer sized~\cite{Togawa2015,Yonemura2017,Clements2017,Han2017,Mito2018} and in bulk specimens~\cite{Miyadai1983,Togawa2013}.\par 

Bulk crystals of $\mathrm{CrNb_3S_6}$ were grown using a chemical vapour transport method~\cite{Miyadai1983, Kousaka2009}. A micrometer sized rectangular specimen was cut from a bulk crystal using a focused ion beam technique and attached onto the signal line of a coplanar waveguide using tungsten. The length along the helical axis, the width parallel to $H$, and thickness of the specimen were 58.6, 12.4 and 2.6~$\mathrm{\mu}$m, respectively.\par
 
A coplanar-waveguide based microwave spectroscopy technique was employed to measure the spin excitation spectra via the forward transmission parameter $S_{21}$ as a function of the frequency and the magnetic field (see Supplemental Material for details). Herein, the corrected magnitude and the field derivative of $S_{21}$ are referred to as $\Delta S$ and d$\Delta$S/d$H$, respectively. In the experiments, the magnitude of the external field, $|H|$, was first decreased from large field values (well above $H_C$) to 0~T and then increased from 0~T up to large fields, and these are referred to as decreasing an increasing field sweeps, respectively. 


\begin{figure}[t]
\centering
\includegraphics[width=8.7cm]{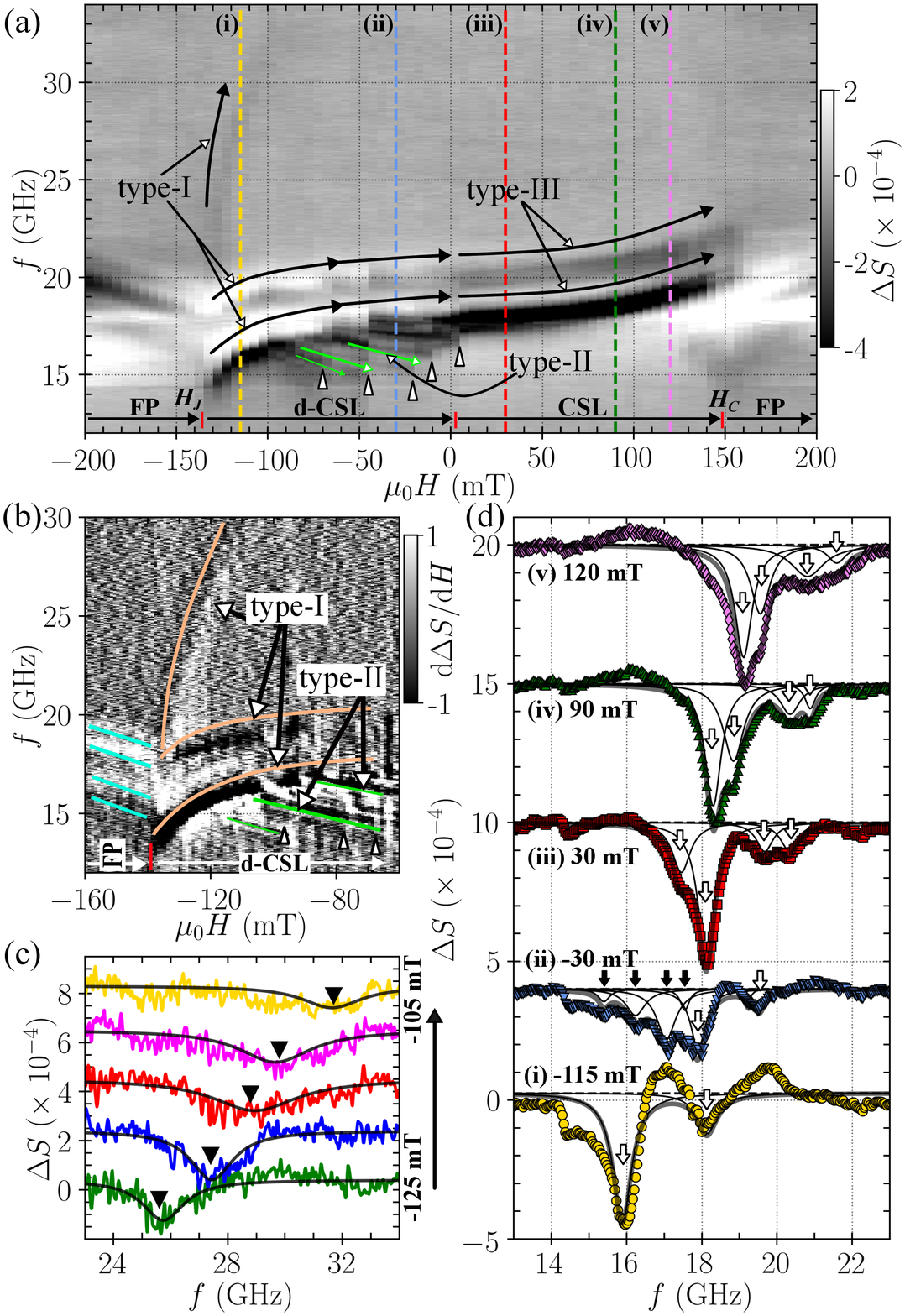}
\colorcaption{(a) Amplitude of $\Delta S$ plotted as a function of frequency, $f$ and external magnetic field, $H$. Data obtained while sweeping $\mu_0H$ from $\mathrm{-200~mT}$ to +200~T in steps of 5~mT, at a temperature of 20~K. The horizontal lines indicate the magnetic phases FP, disordered-CSL, CSL and their arrowheads indicate the field field sweep direction. The elongated triangles displayed vertically indicate the fields where the number of type-II modes changed abruptly. (b) Plot of d$\Delta S$/d$H$ as a function of $f$ and $H$. Data obtained while a sweeping $\mu_0H$ from $\mathrm{-200~mT}$ to +200~mT in field steps of 1~mT. (c) Line trace of the high order mode plotted as a function of $f$ in the vicinity of $H_C$. Markers indicate the center frequency of the resonance. (d) Amplitude profiles of $\Delta S$ at the various stages of the field sweep (dashed lines in (a)). Open and full black arrows indicate the center frequency of the resonance modes of type-I/III and type-II modes, respectively. The line traces are Lorentzian fits to the various resonance modes.}
\label{fig:FIG1}
\end{figure}

Figure~\ref{fig:FIG1}(a) shows the amplitude of $\Delta S$ obtained while varying the magnitude of $\mu_0H$ from $\mathrm{-200~mT}$ to 200~mT. In the field polarized (FP) phase, we identified three to four resonance modes, which exhibited a Kittel-type field dependence. At a magnetic field of -138~mT ($H_J$), the Kittel-like modes were replaced by three resonance branches whose field dependence is highlighted by the curved lines labeled as type-I modes. This transformation in resonance behavior marked the emergence of the CSL phase. As the field was decreased further, the resonance frequency of the two type-I modes at lower frequencies varied between 14 to 20 GHz while the frequency of the third mode increased rapidly from 18~GHz to 35~GHz. Between $\mathrm{-138~mT}$ and $\mathrm{-100~mT}$ the type-I modes followed the outset of a dome-like field dependence, which is consistent with the expected behavior in the intrinsic CSL phase~\cite{Kishine2009,Kishine2015}. In order to clarify the field behavior of the type-I modes, we show the field derivative plot of $\Delta S$ in Fig.~\ref{fig:FIG1}(b) which highlights the existence of three resonance modes. Figure~\ref{fig:FIG1}(c) shows the amplitude profile of the resonance mode observed at high frequencies, for different values of $H$.\par
 
In the field range between $\mathrm{-100~mT}$ and $\mathrm{0~mT}$, in addition to the type-I modes, we observed a number of other resonance modes which clearly do not follow the field dependence intrinsic to the collective dynamics of the CSL. These are indicated in Figs.~\ref{fig:FIG1}(a) and (b) as type-II modes. With decreasing the field magnitude towards 0~T, each of these modes decreased linearly in frequency and disappeared suddenly at certain fields, as indicated by the triangular markers. In this field region we observed five to six resonance modes which vanished sequentially from lower to higher frequencies.\par

The switching of the field polarity, from small negative to positive fields ($\mathrm{-5\rightarrow}$5~mT), which expectedly enabled the appearance of the helical state, resulted in both the vanishing of the type-II modes and the sharp increase in the amplitude of the microwave absorption of the resonance modes identified as type-III in Fig.~\ref{fig:FIG1}(a). The change in the resonance spectra can be visualized also in Fig.~\ref{fig:FIG1}(d) which shows various absorption lines measured before and after 0~T. Note that the frequency spreading of the resonance modes in (ii) is wider by 3~GHz compared to (iii) due to the existence of the type-II modes, indicated by the solid arrows, at lower frequencies.\par 

Despite the much larger resonance amplitude above 0~T, the frequency of the two pairs of type-III modes seemed to connect smoothly across 0~T to the type-I modes observed in the decreasing field branch. However, at increasingly large values of $H$ in the increasing field process, the resonances deviated from the dome-like field dependence of the type-I modes. Instead, the resonance frequency increased slowly at low fields and then more rapidly as $H$ approached $H_C$. Note that in the increasing field branch ($H>$ 0), each broad resonance was comprised of two overlapping modes with similar field dependences, as shown in Fig.~\ref{fig:FIG1}(d). Importantly, the resonance frequency of the type-III modes followed a field dependence that is neither the intrinsic dome-like CSL behavior of the type-I modes nor the Kittel-like dependence of the type-II modes observed in the decreasing field branch.\par  
 
In the resonance spectra observed near $H_C\sim$150~mT we observed a field region, between 140~mT and 160~mT, where the resonance modes attributed to the CSL and FP phases appear to coexist. We chose $H_C$ as the field magnitude above which the resonance attributed to the CSL decreased pronouncedly while the kittel-like resonances have clearly recovered. Well above $H_C$, the resonances exhibited a field dependence similar to that of the Kittel-like modes previously identified as the FP phase, between $\mathrm{-200~mT}$ and $\mathrm{-138~mT}$ .\par

Previous experimental and theoretical results have shown that the collective spin dynamics of the CSL in finite size specimens depends strongly on the shape, the boundary conditions and on the intrinsic dynamics of the MKs~\cite{Kishine2016,Goncalves2017,Goncalves2018}. Particularly in Refs.~\onlinecite{Goncalves2017,Goncalves2018}, which focused on micrometer sized specimens with a reduced number of MKs, the resonance behavior in the decreasing field process was characterized by a frequency jump of typically 1.5-2.0~GHz, at $\mu_0H_J \sim$~0.6$H_C$. This abrupt jump has been widely observed through various physical properties and has been recently attributed to the existence an energy barrier that controls the insertion of the MKs in the decreasing field process~\cite{Shinozaki2018}. In the present specimen, the frequency jump was not clearly observed. The absence of a clear jump in the resonance frequency and the fact that the magnitude of $H_J$ is comparable to $H_C$ appears to have promoted a near continuous formation of the CSL with decreasing $\mu_0H$ from $\mathrm{-138~mT}$ to 0~T. Hence, the observation of a dome-like field dependence on the type-I modes during the field decreasing process.\par 

The dome-like field dependence of the type-I resonance modes is consistent with the previous theoretical studies on the intrinsic excitation spectra of the CSL. Particularly in Ref.~\onlinecite{Kishine2009}, where, in addition to the overall dome-shaped field dependence, it is discussed that the $n$th order CSL modes are dependent on a set of wavevectors linked to the spin modulation period of the MKs and that these modes can exist over a wide frequency range. Moreover, a rapid increase in the resonance frequency of the higher order modes and a simultaneous decrease of the mode amplitude is expected to occur with increasing the density of MKs.\par

\begin{figure}[]
\centering
\includegraphics[width=8.7cm]{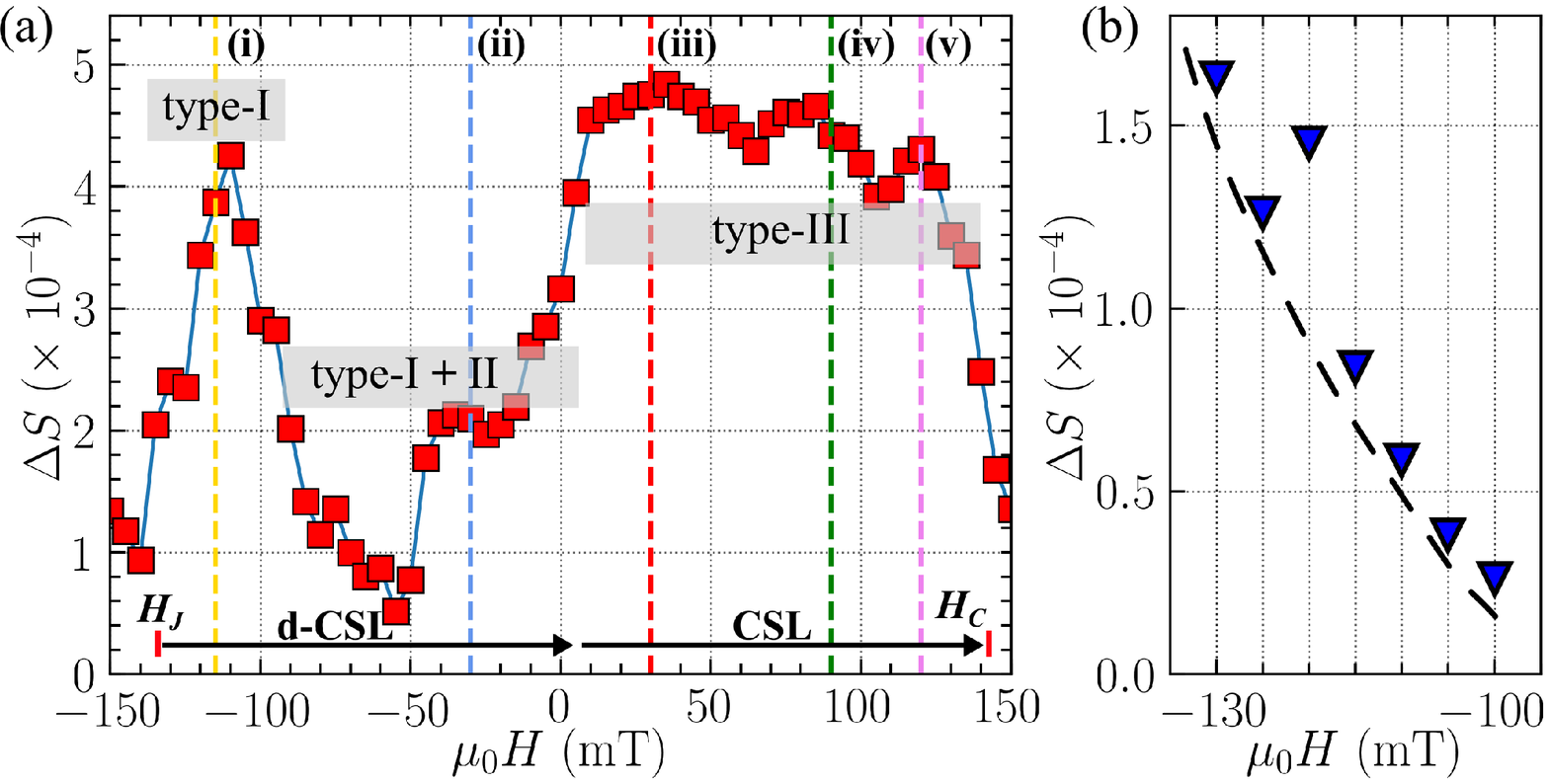}
\colorcaption{(a) Absorption amplitude of the main resonance peak of type-I and type-III modes attributed to the CSL, plotted as a function of $H$. The colored lines mark the same fields as shown in Fig.~\ref{fig:FIG1}. (b) Absorption amplitude of the high order mode as a function of $H$ in the vicinity of $H_J$ (dashed line is a guide to the eye).}
\label{fig:FIG2}
\end{figure}

In the present Letter, we experimentally confirm the magnonic character of the CSL through the characterization of the high order mode which so far has only been discussed from the theoretical standpoint in Ref.~\onlinecite{Kishine2009}. In particular, we confirmed that the collective resonance of the CSL increased markedly, from about 18~GHz up to 35~GHz, and that the amplitude of the high order mode varied rapidly (see Fig.~\ref{fig:FIG2}(b)) in the vicinity of $H_J$ (or $H_C$), where a drastic change in the density of MKs is expected to occur.\par
The field dependence of the resonances is discussed further with reference to Fig.~\ref{fig:FIG2}(a), which shows the absorption amplitude of the main type-I ($H<$~0) and type-III ($H>$~0) modes plotted as a function of $H$. We observed an increase in the amplitude of the main type-I mode between -138~mT and -100~mT, followed by a decrease at $\mu_0H>$ $\mathrm{-100~mT}$, ascribed to the emergence of the type-II modes. Above 0~T, a drastic increase in the amplitude of the type-III modes was observed.\par 
The coexistence of the low order type-I modes with the Kittel-like type-II modes strongly suggests that the CSL is mixed with regions where ferromagnetic alignment persists, possibly along with a large number of magnetic dislocations. The number and amplitude of type-II modes in the decreasing field process is linked to the spatial distribution and extent of the magnetic dislocations which in turn reflects on the contribution from the regions with ferromagnetic alignment. Presumably, such overall disordered state was responsible for the partial disruption of the intrinsic collective resonance of the CSL, thereby preventing the CSL from acting as a collective entity at a macroscopic length scale.\par
The enhancement of the collective resonance behavior of the type-III  modes ($H>$~0~T) is believed to have been enabled by the expulsion of the magnetic dislocations. \par

So far, only spectroscopic techniques such as microwave resonance have been employed in the detection of the disordered CSL phase mainly due to specimen thickness (2.6~$\mathrm{\mu m}$). However, recent studies where the magnetic texture was imaged in much thinner specimens, using the Lorentz mode of transmission electron microscopy, showed the existence of magnetic edge dislocations, which are a signature of magnetic disorder in the CSL phase. It is important to note that, in a clear resemblance to the data discussed here, the magnetic dislocations were only observed while imaging in the decreasing field process. These results will be reported elsewhere~\cite{GPaterson2019}.\par

With changing the polarity of $H$ via zero magnetic field and the consequent emergence of the chiral helical state, the type-II modes vanished, triggering a different form of collective dynamics of the CSL. This is clearly seen through the increase in amplitude of the type-III modes above 0~T and the sloped field dependence which differs from that observed in the type-I modes. In fact, a behaviour similar to that of the type-III modes has been seen in previous experiments~\cite{Goncalves2017}, where a sloped field dependence was observed when following the same excitation configuration. Presently, the mechanism behind this sloped field behavior is not clearly understood. More theoretical and experimental work is necessary on this front.\par 

In the light of the present experiments, one possible explanation might be that the collective resonance of the CSL became strongly dependent on the boundary spins due to the re-ordering of the spins near the surface (at 0~T). While the type-I modes reflect the intrinsic response of the CSL, the type-III modes are a result of a collective CSL dynamics imposed by the boundary conditions which ultimately are dependent on the strength of the external field (hence the sloped behavior). The absence of ferromagnetic domains and magnetic dislocations contributed to an increase in the absorption amplitude and an enhancement in the coupling between the microwave fields and the collective spin excitation modes of the CSL defined by the boundary spins. In the decreasing field process, the effect of the boundary spins on the collective CSL dynamics could have been suppressed due to the existence of a disordered phase.\par 

\begin{figure}[t]
\centering
\includegraphics[width=8.7cm]{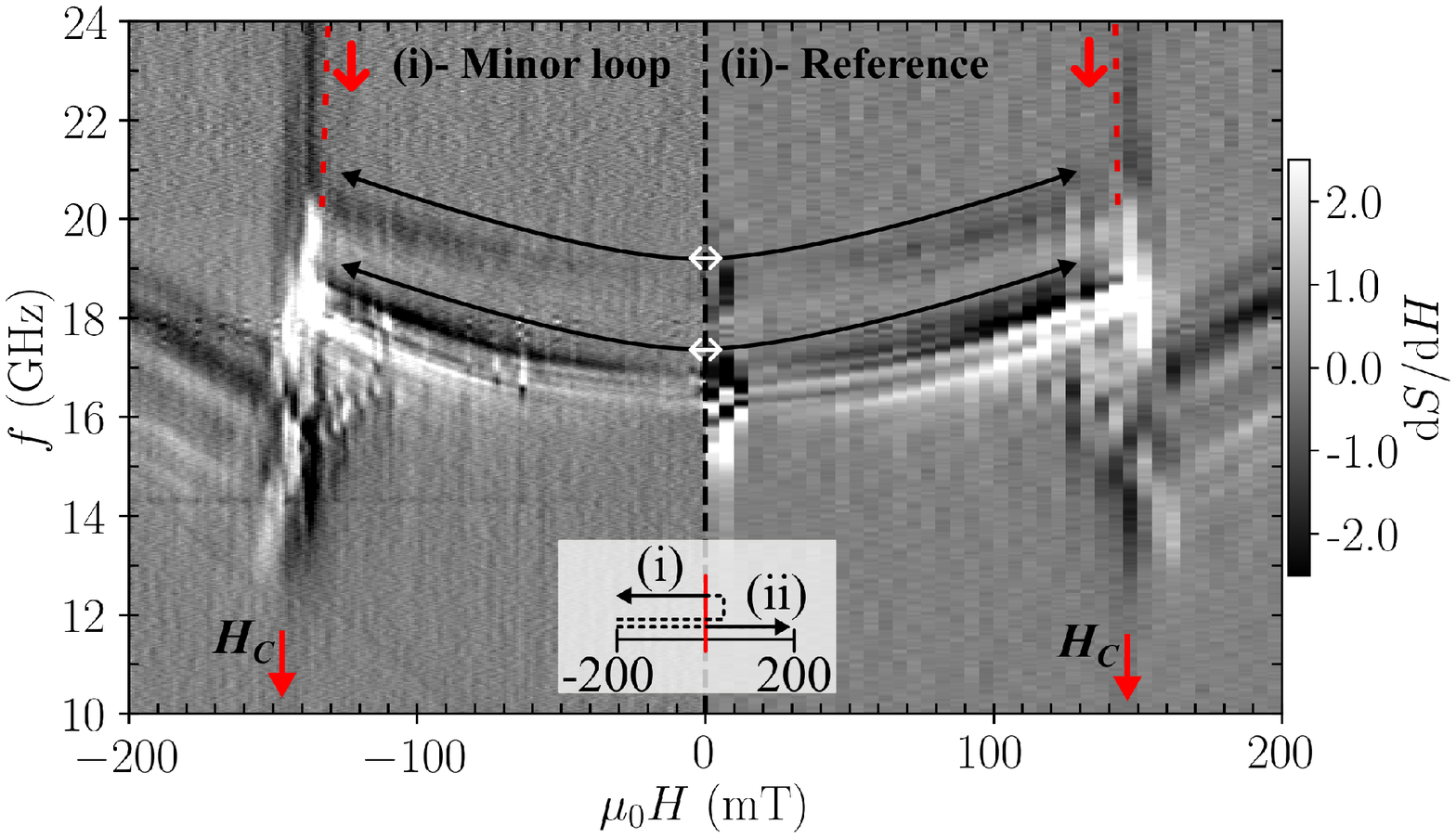}
\colorcaption{Plot of d$\Delta S$/d$H$ as a function of $f$ and $H$, corresponding to the increasing field branch of (i) a minor field loop and (ii) the increasing field branch obtained in a field sweep from $\mathrm{-200~mT}$ to 200~T. Data obtained at a constant temperature of 50~K. The horizontal arrows indicate the field sweep direction and inset illustrates the field process followed in (i) and (ii).}
\label{fig:FIG3}
\end{figure}

Importantly, it is experimentally demonstrated that the disordered magnetic phase can be switched off and the collective dynamics restored if the specimen enters the helical phase, concurrent with the change in the polarity of the magnetic field. In Fig.~\ref{fig:FIG3}, we demonstrate the applicability and robustness of the resonance spectra corresponding to a magnetically ordered CSL phase at a macroscopic length scale. Once the disorder was removed from the specimen, the enhanced collective resonance persisted regardless of which field sweep direction is adopted (0 $\rightarrow$ $\mathrm{-200~mT}$ in (i) or 0 $\rightarrow$ 200~mT in (ii)). Clearly, this enhanced response is symmetric with regards to the magnitude of $H$, which contrasts with the disrupted CSL dynamics obtained when in the presence of magnetic disorder, as already presented in Fig.~\ref{fig:FIG1}. \par 

In the data shown in Fig.~\ref{fig:FIG1}(a), the high order mode is not clearly identified in the increasing field branch due to loss in sensitivity in that experiment. However, the data presented in Fig.~\ref{fig:FIG3} clearly points towards the existence of the high order resonance mode, as indicated by the red vertical arrows and dashed lines both in (i) and (ii) (see also Fig.~2 in the Supplemental Material).\par  

Following the interpretation above, the experimental results may be summarized in the following manner. A disordered CSL, comprised of CSL and embedded ferromagnetic domains, produced two types of resonance modes with distinct and independent field behavior. As $H$ approached 0~T, the resonance attributed to the ferromagnetic domains disappeared gradually until the helical phase was reached at 0~T. At this point, a pronounced increase in the amplitude of the resonance modes was observed, suggesting that the ferromagnetic domains were disrupting the collective spin precession of the CSL. Once the helical phase was reached, the whole system resumed the collective behavior, which consisted of a number resonance modes with large amplitude and similar field behavior. Thus, we have demonstrated that the disordered state can be completely erased simply by sweeping the external field to take the sample through the helical magnetic phase at 0~T.\par 

The control over the degree of magnetic disorder in the CSL, which was facilitated by choice of field process, can potentially enable the formation of field controlled channels of MKs that could serve as SW conduits. Further control over magnetic disorder may be achieved via the choice of specimen dimensions, within the sub-mm range~\cite{Goncalves2018}, and modifications to the surfaces of the specimens.

FG and YT would like to thank T. Yuki for the support in the sample fabrication, A. Ovchinnikov, J. Kishine and I. Proskurin for the fruitful discussions on the manuscript and K. Inoue and J. Akimitsu for promoting discussions around the topic of chirality in condensed matter. The authors acknowledge the support from Grant-in-Aid for Scientific Research (Nrs. 25220803, 17H02767, 17H02923), Chirality Research Center (Crescent) in Hiroshima University, JSPS, RFBR under the Japan - Russia Research Cooperative Program, JSPS Core-to-Core Program, A. Advanced Research Networks. FG received support from JSPS International Research Fellowship No. 17F17316 and GWP received support from EPSRC Grant No. EP/M024423/1.

\bibliographystyle{apsrev4-1}
\bibliography{ms}

\clearpage
\begin{center}
\textbf{\large Supplemental Material \\ Effect of disorder on the collective excitations of the chiral spin soliton lattice}
\end{center}

The supplemental material contains two sections. In Sec.~\ref{sec:one} we provide details on the experimental procedure. In Sec.~\ref{sec:two} we discuss the origin and field robustness of the type-I and type-II modes presented on the manuscript. In particular, we discuss experimental data concerning a minor field loop that helped identifying the origin of the type-II modes. Additionally, we provide experimental evidence for the observation of the higher order type-I modes on both the decreasing and increasing magnetic field branches.

\setcounter{figure}{0}
\renewcommand{\thefigure}{S\arabic{figure}}
\section{Experimental procedure}\label{sec:one}
\begin{figure}[]
\centering
\includegraphics[width=6.0cm]{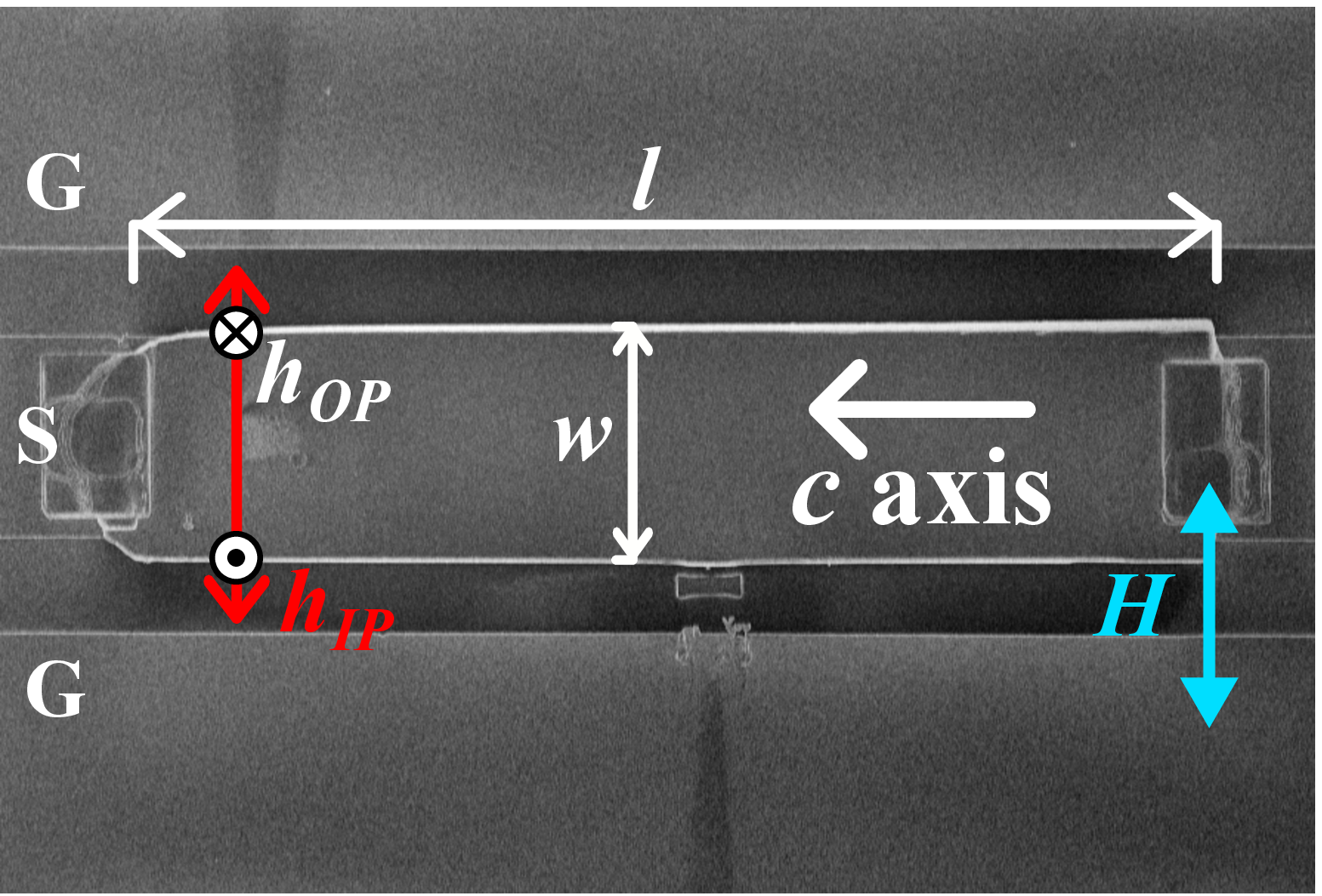}
\colorcaption{Scanning electron microscopy image of the coplanar waveguide and the specimen placed on top of its signal line (S). The length ($l$), width ($w$) and thickness of the specimen were 56.8~$\mu m$, 12.4~$\mu m$ and 2.6~$\mu m$, respectively. The arrows indicate the orientation of the external field and the microwave field components relative to the helical (c)axis of the crystal.}
\label{fig:FIG1sup}
\end{figure}

Broadband frequency coplanar waveguide type microwave resonance spectroscopy was employed to measure the resonance response of the micrometer sized crystal. A microwave current flowing in the signal line, supplied and analyzed by a vector network analyser via the forward transmission parameter $S_{21}$, induced transverse in-plane ($h_{IP}$) and out-of-plane ($h_{OP}$) microwave field components which acted on the magnetisation precession in the specimen. The resonance data was extracted from the forward transmission parameter, $S_{21}$, measured in a vector network analyser as a function of the microwave frequency ($f$), which was swept between 0.1-40~GHz, in the presence of an external magnetic field. A reference spectra was subtracted from the experimental data in order to remove unwanted non-magnetic artifacts related to circuit loss. As a result of the position of the specimen with regards to the waveguide, the in-plane and out-of-plane components of the microwave field were collinear with the width and the thickness of the specimen, respectively while perpendicular to the helical axis. The orientation of the helical or $c$-axis was set perpendicular to both the external magnetic field, $H$, and the microwave field components $h_{IP}$ and $h_{OP}$, as illustrated in Fig.~\ref{fig:FIG1sup}.\par
The orientation of the helical axis relative to the static external field resembles configuration (II) examined in Ref.~\cite{Goncalves2017}, where the resonance signals attributed to the field polarised and the CSL phases were observed within a field magnitude of 200~mT. The quantity $S_{21}$ was acquired as a function of frequency from 1 to 40~GHz and magnetic field, from 200~mT to $\mathrm{-200~mT}$. A reference spectra was used to subtract the background signal in order to minimise the effect of non-magnetic features unrelated to the resonance signal of the specimens (predominantly circuit loss). \par 
The corrected transmission spectra and its field derivative are referred to as $\Delta S$ and d$\Delta S$/d$H$, respectively. The experimental data presented and shown in Figs.~1-2 of the manuscript were acquired at 20~K, while the data corresponding to the minor field loops, shown in Fig.~3 of the manuscript and Fig.\ref{fig:FIG2sup}, were acquired at 50~K. The values of the resonance frequencies are lower compared to the data acquired at 20~K (Figs.~1-2) due to the temperature dependence of the easy-plane anisotropy~\cite{Miyadai1983}. Nevertheless, the field dependence was generally the same at 20~K and at 50~K since both temperatures are well bellow the transition temperature of 128~K.\par

\section{Minor field loop}\label{sec:two}

The degree of functionality that can be achieved with controlling the magnetic disorder in the specimen, was demonstrated in the data corresponding to the minor field loop discussed with Fig.~3 of the manuscript, where we highlighted the importance of the transformation brought by the helical state (0~T).\par
Further evidence for the origin of the type-I and type-II modes was obtained by measuring the resonance during a field sweep in which the helical phase has not been reached. The results showed that the type-I modes clearly underwent the phase transition at $H_J$ while the type-II modes, despite sensitive to the magnetic phase transition, connected to the FP phase and are symmetric around the reversing field point. The linear field dependence of the type-II modes is consistent with Kittel-like modes typically observed in conventional ferromagnetic elements, showing unambiguously that the origin of the type-II modes is linked to the existence of small domains comprised of field polarised spins and other features such as magnetic dislocations which promote the emergence of collinear spins embedded in the CSL phase. In this disordered CSL phase, the type-I and type-II modes behave independently, meaning that each mode type has its own field dependence.\par
\begin{figure}[]
\centering
\includegraphics[width=8.7cm]{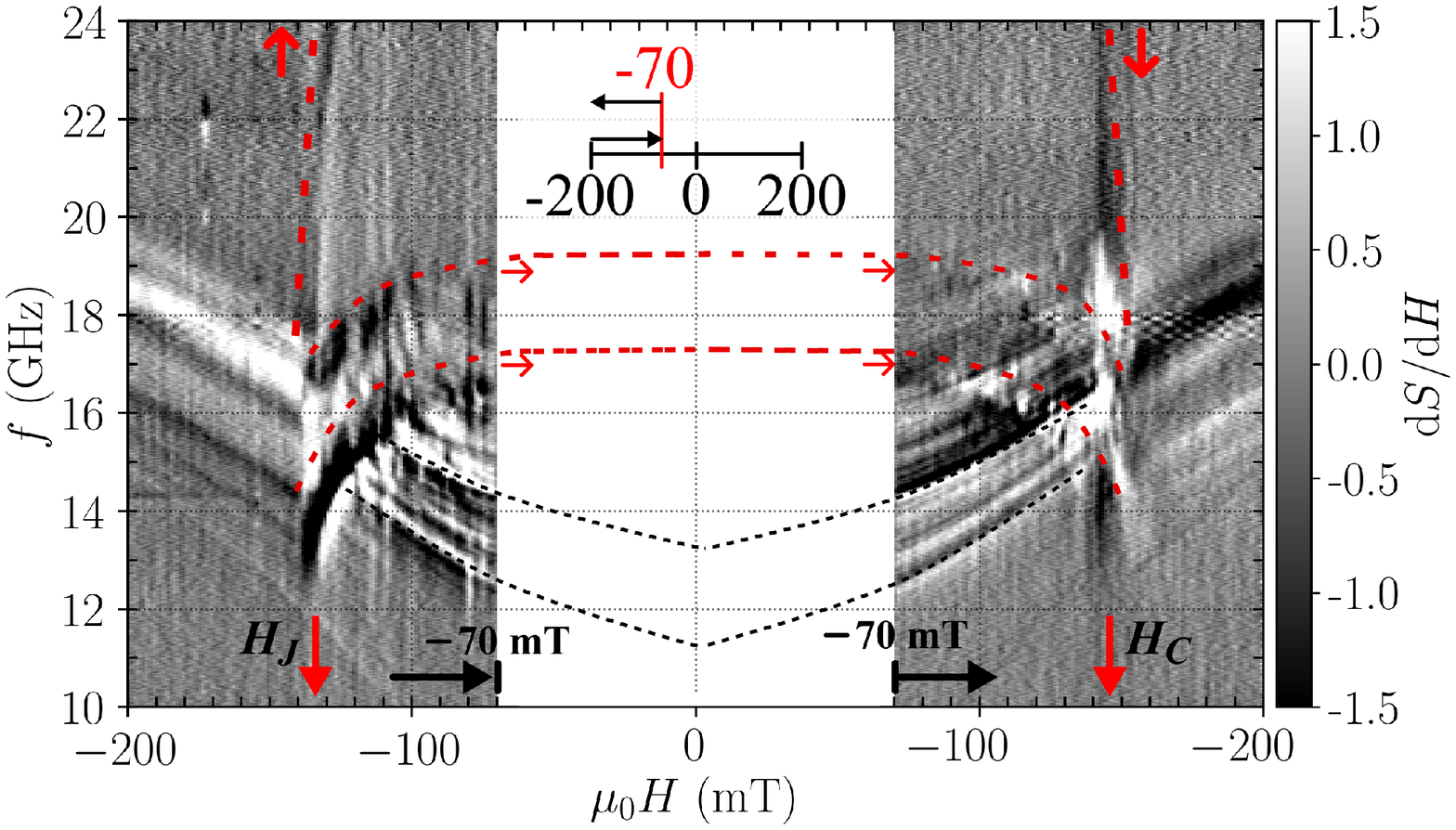}
\colorcaption{Quantity d$\Delta$S/d$H$, plotted as a function of $f$ and $H$, corresponding to the minor field loop illustrated on the inset schematic. Data obtained at a constant temperature of 50~K. The horizontal arrows indicate the field sweep direction.}
\label{fig:FIG2sup}
\end{figure}

In the minor loop shown in Fig.~\ref{fig:FIG2sup}, the external field was varied from $\mathrm{-200~mT}$ to $\mathrm{-70~mT}$ followed by a reverse from $\mathrm{-70~mT}$ to $\mathrm{-200~mT}$. In the decreasing field process shown on the left, the specimen underwent the transition into the CSL phase at $\mathrm{-138~mT}$. Bellow this field value, the type-I modes (indicated by the red dashed lines) and the type-II resonance modes have equal predominance and yet distinct field dependence. It is apparent that some of the type-II modes have the same field slope as that of the resonances obtained in the FP phase. On the other hand, it is also clear that other type-II modes branch off the type-I modes, particularly those at higher frequencies. \par In the increasing field branch, shown on the right, the slope of the type-II modes changes sign and near $H_C$, some of the type-II modes connect smoothly to the FP phase. In contrast, the type-II modes observed at higher frequencies vanish when $H_C$ is reached. Here, we also observed evidence for the coexistence of the type-I modes, which appear to exhibit the dome-like field behaviour on the two modes highlighted by the red dashed lines. Moreover, although limited by the upper frequency limit used in these measurements (24~GHz), we note hints of the third, higher order mode in these measurements on both the decreasing and increasing field branches of this minor field loop.\par 
 
In summary, the minor field loop presented in Fig.~3 of the manuscript demonstrated that when starting from the ordered CSL phase, imposed only by reaching the helical state, any field process will resulted in collective resonance modes which undergo similar (symmetric) field evolution. However, as demonstrated in Fig.~\ref{fig:FIG2sup} if the starting point was any other than the helical state, the field process that the type-I and type-II modes behaved independently. This way, we further demonstrate the degree of functionality that can be achieved with controlling the magnetic disorder in the specimen.\par

\end{document}